\documentclass[prl,aps]{revtex4}
\usepackage{graphicx}%
\usepackage{amsmath}

\begin{document}
\title{Superluminal behavior and the Minkowski space-time}
\author{Daniela Mugnai\footnote{E-mail: d.mugnai@ifac.cnr.it}}
\affiliation{``Nello Carrara'' Institute of Applied Physics,  CNR
Florence Research Area,\\
 Via Madonna del Piano 10, 50019 Sesto
Fiorentino (FI), Italy}

%\pacs{42.25.Bs, 03.30.+p}

\begin{abstract}
\vspace{1 cm}

 Bessel X-waves, or Bessel beams,  have been
extensively studied in last years, especially with regard to the
topic of superluminality in the propagation of a signal. However,
in spite of  many efforts devoted to this subject, no definite
answer has been found, mainly for lack  of an exact definition of
signal velocity. The purpose of the present work is to investigate
the field of existence of Bessel beams in order to overcome the
specific question related to the definition of signal velocity.
Quite surprisingly, this field of existence can be represented in
the Minkowski space-time by a Super-Light Cone which wraps itself
around the well-known Light Cone.
% So, the change in the upper limit of the light velocity does
%not modify the fundamental low of the relativity and the causal
%principle.
\end{abstract}

\maketitle

\newcommand{\be}{\begin{equation}}
\newcommand{\ee}{\end{equation}}
\newcommand{\bea}{\begin{eqnarray}}
\newcommand{\eea}{\end{eqnarray}}

The propagation of Bessel X-waves has been extensively analyzed in
last years, especially with regard to the topic of superluminality
in connection to the signal propagation. Many contributions were
devoted to this topic, both from a theoretical and experimental
point of view \cite{saa, mug00,ale,bes,zam,saa1}.

Bessel X-waves, which are also known as Bessel beams, belong to
the class of localized waves. The peculiarity of this type of
waves is that they are well localized in space, unlike a ``usual"
wave which occupies the entire space. As is well known, a $u_B$
Bessel beam  is the result of superimposing an infinite number of
plane waves, each of them with a direction of propagation tilted
by the same angle $\theta_0$ with respect to a given axis, say z.
In cylindrical coordinates ($\rho,\, z,\psi$), the beam is given
by

 \be
  u_B(\rho ,z,t) = J_0(k_0\rho\sin\theta_0 )\exp \left[i k_0 z \cos\theta_0
\right]\,\exp(-i\omega t)
 \label{bes}
 \ee
 where $k_0=\omega /c$ is the wavenumber in the vacuum, and
$\omega$ is the frequency of the beam. Function  $J_0$ denotes the
zero-order Bessel function of first kind, which, apart from
inessential factors, can be written as \cite{abr}

\be
J_0(x)=\int_0^{\pi} \exp (ix\cos\varphi )\,d\varphi .
 \label{delta}
 \ee
The characteristic  features of a Bessel beam are that it supplies
well-localized energy,  that  propagates along the z-axis with no
deformation in its amplitude \cite{dur,spr}, and that both phase
and group velocities are greater than the light velocity $c$
\cite{mug00,ale}.

A $U_B$ Bessel pulse limited in time, which is the theoretical
definition of signal, can be obtained by superimposing  an
infinite number of frequencies. After integration of Eq.
(\ref{bes}) over $d\omega$,  and by substituting the Bessel
function $J_0$ with its integral form,  we obtain

\be
 U_B (\rho, z, t)= \int_0^\pi\delta\left(\frac{\rho}{c}\sin
 \theta_0 \cos\varphi +\frac{z}{c} \sin\theta_0
 -t\right)\,d\varphi\,,
 \ee
which is different from zero only if
\be
t \leq \left| \frac{1}{c} \left( z\cos\theta_0
 +\rho\sin\theta_0\right)\right|,
  \label{tb}
\ee
 where $0\leq\theta_0<\theta_{max}$, $\theta_{max} \ll \pi /2$
 depends on the experimental set-up.

 \begin{figure}
\includegraphics[width=0.5\textwidth]{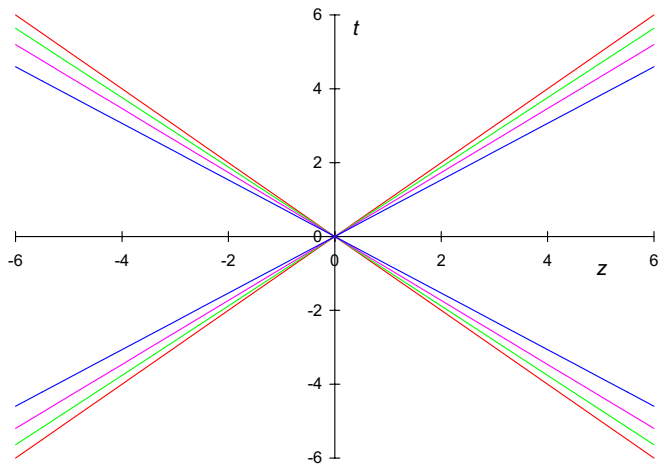}
  \caption{{\small Bessel beam velocities for
three different values of parameter $\theta_0$, in the $z-t$
plane, for $\rho=0$ (Euclidean space) . The red line indicates
light velocity $c$ (for $\theta_0=0$) taken here as equal to 1.
Green, pink, and blue lines represent the beam velocities for
$\theta_0 = 20^\circ,\, 30^\circ$, and $4^0\circ$, respectively.
For $\rho\neq 0$, the intersection point changes its position with
no modification in the line behavior.}}
\end{figure}

Thus, the time interval in which the beam is different from zero
is
\be
 t_{min}(\theta_0=\theta_{max}) \leq t<
 t_{max}(\theta_0=0).
 \ee

 \begin{figure}
\includegraphics[width=0.4\textwidth]{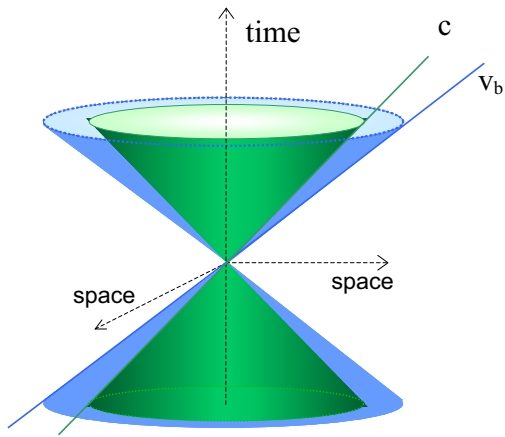}
\caption{{\small Schematic representation of the Super-Light Cone
in the Minkowski space-time (pseudo-Euclidean space). The green
zone represents the Light-Cone, while the blue zone around it is
the field of existence of the Bessel beam. Quantity $v_b$ is the
beam velocity for a given axicon angle, $\theta_0$. For
$\theta_0=0$, the beam is reduced to a plane wave, and its
velocity then becomes equal to $c$. In this situation,  the field
of existence of the beam goes to zero, the Super-Light Cone
narrows and becomes equal to the Light-Cone.}}
\end{figure}

 Since the Bessel pulse propagates along the z-axis,  we can deduce  that the
motion of the beam in the $z-t$  plane (see Fig. 1) is within a
conical surface similar to the Light Cone, where light velocity
$c$ is replaced by velocity  $v_b = c/\cos\theta_0$, and $t$ is a
real quantity: we can say that the propagation of a Bessel pulse
in the Euclidean-space corresponds to a Super-Light Cone in the
pseudo-Euclidean space-time of  Minkowski. In other words, by
introducing a second spatial coordinate, for a given value of
$\theta_0$, we obtain a Super-Light Cone like the one of Fig. 2,
where straight line $v_b$, which depends on $\theta_0$, is the
beam velocity.

For $\theta_0=\theta_{max},\: v_b$ represents the border line
which determines the existence of the field: the Bessel beam
exists only in the blue zone. Inside this cone of existence, the
past Super-Light Cone, $t < 0$, represents the time interval prior
to generation of  the beam. The beam originates at $t=0$, and for
$t>0$ (future Super-Light Cone) propagates along the $z$ axis with
velocity  $v_b$ (blue line, in Fig. 2). For $\theta_0=0$  the beam
reduces to a plane wave, its velocity becomes equal to $c$ (green
line, in Fig. 2), and the Super-Light Cone becomes the Light Cone
(green cone in Fig. 2).

Now, since Bessel beams are real quantities  (they have been
experimentally generated and measured), and since Eq. (\ref{bes})
is capable of describing the scalar field of the beam as being due
to a specific experimental set-up \cite{mug06},  we can conclude
that the Super-Light Cone places a new upper speed limit for all
objects. Massless particles  can travel not only along the Light
Cone, but also along the Super-Light Cone in the region between
the Super-Cone and the Cone, while the world-lines remain confined
within the Light-Cone.  In substance, we can think that $c$ is the
velocity of  light in its simplest manifestation (wave), while
more complex electromagnetic phenomena, such as the interference
among an infinite number of waves, may originate different
velocities. The maximum value $\theta_{max}$ of  axicon angle
$\theta_0$ sets the maximum value of the beam velocity.  Since the
filed depth, that is, the spatial range in which the beam exists,
is proportional to $\tan\theta_0^{-1},\: \theta_0$  can never
reach the value of $\pi /2$. If it were possible to obtain values
of $\theta_0$ close to $\pi /2$, we should have almost  immediate
propagation in a nearly-zero space, rather like an ultra fast shot
destined to slow down immediately.

The change in the upper limit of the light velocity (the Bessel
beam is ``light") does not modify the fundamental principles of
relativity and the principle of causality, as demonstrated by
recent theory dealing with new geometrical structure of space-time
\cite{car}.  The principle that  ``the speed of light is the same
for all inertial observers, regardless of the motion of the
source", remains unchanged, provided that the substitution
$c\longrightarrow v_b\, ( = c / \cos\theta_0 )$ is made in the
Lorentz transformations. In this way, the direction of the
beam-light does not depend on the motion of the source, and all
observers measure the same speed ($v_b$) in all directions,
independently of their motions.

%\vspace{1 cm}
%\begin{acknowledgments}
%Thanks are due to Laura Ronchi Abbozzo and Roberto Mignani for
%useful discussions.
%\end{acknowledgments}

\end{document}